\documentclass[aps,pre,reprint]{revtex4-2}

\usepackage{amsmath}
\usepackage{amssymb}
\usepackage{amsfonts}

\usepackage{graphicx}
\usepackage{color}
\usepackage{bm}
\usepackage{dcolumn}

\newcommand{\diff}{\mathop{}\!\mathrm{d}}

\newcommand{\vth}{\ensuremath{v_\textrm{th}}}
\newcommand{\vosc}{\ensuremath{v_\textrm{osc}}}

\newcommand{\kB}{k_\mathrm{B}}

\newcommand{\maxw}{Maxwellian\,}
\newcommand{\sg}{Supergaussian\,}

\newcommand{\co}{(color online) }

\newcommand{\varparallel}{\mathbin{\!/\mkern-5mu/\!}}

\begin{document}

\title{Classical molecular dynamic simulation to assess the non-Maxwellian behavior of inverse bremsstrahlung heating in weakly coupled plasmas}
\author{\surname{R.} Devriendt}
\affiliation{CEA, DAM, DIF, F-91297 Arpajon, France}
\author{\surname{O.} Poujade}\email{olivier.poujade@cea.fr (corresponding author)}
\affiliation{CEA, DAM, DIF, F-91297 Arpajon, France}
\affiliation{Universit\'e Paris-Saclay, CEA, LMCE, F-91128, Bruy\`eres-le-Ch\^atel, France}

\date{\today}

\begin{abstract}
	
	Classical molecular-dynamics simulations (CMDS) have been conducted to investigate the non-\maxw behavior of a weakly coupled plasma submitted to the inverse bremsstrahlung heating of laser irradiation. It is the so called {\it Langdon effect}. It has important consequences on plasma properties. It reduces laser absorption, it modifies atomic physics due to its influence on the free electron population, it alters conduction and any other quantities that depends upon electron velocity distribution (EVD). Here, the Langdon effect has been studied with CMDS using the code LAMMPS where, contrary to Fokker-Plank simulations, widely used in the past, plasma many-body behavior at the microscopic level is taken into account self-consistently. For the first time with CMDS, we have observed the deformation of the instantaneous EVD in Langdon's condition. Anisotropy of these non-\maxw effects have been demonstrated at moderate and high intensities. CMDS results (related to the shape of the EVD and to the laser absorption reduction) do not match with previous Fokker-Planck simulations results as well as might have been expected. This should  probably stimulate new developments to understand the complex many-body interaction of electrons and ions in a laser irradiated plasma in the future.

\end{abstract}

\maketitle

\section{Introduction}
Collisional absorption of laser radiation by a plasma through the process of inverse bremsstrahlung (IB) is the dominant absorption mechanism at intensities less than $10^{15}$ W/cm$^2$. A precise modeling of this physical effect in radiation-hydrodynamics codes \citep{troll2018, marinak2001} is paramount to designing high energy density (HED) experiments and inertial confinement fusion experiments \citep{betti2016} in order to reach the goal of thermonuclear ignition \citep{lawson22}.

When a monochromatic electromagnetic wave (laser) of pulsation $\omega$ and local intensity $I$ propagates through a plasma of electronic density $n_e$ and temperature $T_e$, electrons undergo an oscillating coherent motion, at the same pulsation, superimposed to their random thermal motion of characteristic velocity $\vth=\sqrt{\kB T_e/m_e}$. The velocity amplitude of this collective motion is given by $\vosc = e \, E /m_e \, \omega$ known as the quiver velocity where $E = \sqrt{2 \, I/c \, \varepsilon_0}$ is the amplitude of the electric field of the wave. Then, some of the energy carried by the electromagnetic wave is transferred to electrons and ions thermal energies (random motion kinetic energy) by means of their random scatterings. Obviously, this process affects the electron velocity distribution (EVD). It modifies the width of its \maxw shape, since the electron temperature increases, but it can also modify the overall shape, turning the \maxw into a drastically different distribution, such as a \sg as predicted theoretically by Langdon\cite{langdon1980}.

This happens when electron-electron (e-e) collisions are not efficient enough to damp any deviations from the \maxw. Langdon \cite{langdon1980} devised a nondimensional \mbox{parameter} 
\begin{equation}
	\label{eq:dist_defAlpha}
	\alpha = Z \, \dfrac{\vosc^2}{\vth^2} = Z \, \dfrac{2 \, e^2 \, I}{c \, \varepsilon_0 \, m_e \, \omega^2 \, \kB T_e},
\end{equation} such that e-e collisions are negligible when $\alpha \gg 1$. Solving Boltzmann equation with a Fokker-Planck collision term, assuming $(\vosc/\vth)^2 \ll 1$ and neglecting the oscillating component of the e-e collisions (therefore assuming $Z\gg 1$ because of (\ref{eq:dist_defAlpha})), Langdon showed that the isotropic part of the EVD of such a plasma submitted to an oscillating electric field, reaches an asymptotic self similar state described by a \sg  EVD, 
\begin{align}
	f_e^{0}(v) \propto \exp (-A \,v^k)\, ,
\end{align} as will be detailed in \S\ref{sec:dist_tail} of this paper. The parameter $k$ is called the order of the \sg and it tends asymptotically towards a value, depending upon $\alpha$, comprised between $k=2$ (when $\alpha=0$) and $k=5$ (when $\alpha\gg 1$) according to Langdon's derivation. This deformation may result in a reduction of absorption by a significant factor when compared to the case $\alpha=0$ (still assuming $(\vosc/\vth)^2 \ll 1$). Subsequently, Fokker-Planck simulations (FPS) confirmed \citep{jones1982, weng2006} Langdon's finding, and the authors of these simulations proposed \citep{matte1988, weng2009} a fit, with respect to $\alpha$, for the order of the EVD and for the reduction in absorption.

Other shapes of EVDs have been suggested in the literature. A sum of a \sg with a \maxw tail \citep{fourkal2001} was put forward for low laser intensities because e-e collisions could become more important than e-i collisions for the population of electrons lying in the supra-thermal tail of the EVD. As an other example, for high laser intensity ($\vosc/\vth \gg 1$), it is expected \citep{ferrante2001} that temperatures parallel and perpendicular to the electric field polarization grow at different rates. This could result in two-temperature oscillating \maxw \citep{chichkov1992, pfalzner1992} or \sg \citep{porshnev1996} EVDs.

Recent, high energy density experiments confirmed \citep{liu1994, milder2020, turnbull2020, milder2021} that non-\maxw distributions do form in laser induced plasmas. Observations of these EVDs have proved that shapes are clearly non-\maxw and look a lot like supergaussians \citep{langdon1980} but they are not precise enough to discriminate all shapes proposed in the theoretical literature.

In classical molecular dynamics simulations (CMDS), which are numerical simulations at the microscopic scale, EVD distortion was also observed by \citet{pfalzner1998} for $Z=1$. Nevertheless, that distortion was not observed in Langdon's conditions requiring \citep{langdon1980} both $(\vosc/\vth)^2 \ll 1$ and $Z \, \vosc^2/\vth^2 \gg 1$ which can only be fulfilled if $Z\gg 1$. It was observed at $\vosc/\vth \gg 1$, which is the only possibility to get $\alpha\gg 1$ with a $Z=1$ plasma.

\citet{ersfeld2000} studied the anisotropy of the instantaneous distribution at high laser intensity using a Fokker-Planck code (FP) with high-order Legendre polynomials. The isotropic, cycle averaged and exact distribution yielded similar heating rate. However, in Fokker-Planck codes, collisions have to be modelled and an analytical expression of the coulomb logarithm, as a function of plasma state parameters (electronic density, temperature, etc.), has to be provided as an input data. This is not the case in CMDS where collisions and many-body effects are inherently taken into account.

In this work, we will study the distortion of the instantaneous distribution at moderate ionization ($Z=10$ specifically) using classical molecular dynamic simulations in order to probe Langdon's theory for a large range of $\vosc/\vth$, and the effect of this distortion on the electron heating rate. In \S\ref{sec:dist_setup}, we will first start by detailing the setup of our CMDS simulations. In Devriendt-Poujade\cite{devriendt2022}, a similar study was described for $Z=1$ weakly coupled plasmas. The absorption model developed on this occasion will be used in the present paper and will be referred to as the parametrized \maxw absorption model (PMAM). In \S\ref{sec::langdon}, CMDS results for plasmas with $(\vosc/\vth)^2 \ll 1$ and $Z \vosc^2/\vth^2 \gg 1$, which are the exact conditions to observe non-\maxw EVDs according to Langdon theory, will be presented. In \S\ref{sec:dist_highI}, we will then show CMDS at higher laser intensity, $\vosc/\vth > 1$, in order to extend Langdon's results. In \S\ref{sec:dist_tail}, we will then reproduce recent experimental results \citep{milder2021}, where EVDs were found to be the sum of a \sg and a \maxw. Finally, in \S\ref{sec:dist_order}, we will compare the order of the distribution as a function of $\vosc/\vth$ from our CMDS simulations with the predictions of \citet{matte1988}.

\section{Classical molecular dynamic simulations setup}
\label{sec:dist_setup}

We carried out classical molecular dynamic simulations (CMDS) of a two components plasma (electrons and one kind of ions with ionization $Z=10$) using the code LAMMPS \citep{LAMMPS}. Large simulations with $4\cdot 10^5$ electrons and $4\cdot 10^4$ ions where carried out in our study contrasting with $10^4$ particles in \citet{pfalzner1998} or in \citet{david2004}. In CMDS, particles are described by their charges, positions and velocities and evolve according to Newton's law. Electrostatic Coulomb forces between each individual particles (electron-electron, ion-ion and electron-ion) are taken into account and for the problem at hand, namely radiation absorption by inverse bremsstrahlung, an external oscillating electric field, and its interaction with each charges of the simulated plasma, is taken into account. Therefore, the exact equations of motion for electrons and ions in LAMMPS are 
\begin{widetext}
	\begin{align}
		m_e\,\frac{\mathrm{d}\bm{v}_\mathrm{e}^{(\alpha)}(t)}{\mathrm{d}t}&=-\frac{e^2}{4\pi\varepsilon_0}\sum_{\beta\neq\alpha}\frac{\bm{n}_{\alpha\beta}}{|\bm{r}_\mathrm{e}^{(\alpha)}-\bm{r}_\mathrm{e}^{(\beta)}|^2}+\frac{Z\,e^2}{4\pi\varepsilon_0}\sum_{b}\frac{\bm{n}_{\alpha b}}{|\bm{r}_\mathrm{e}^{(\alpha)}-\bm{r}_\mathrm{i}^{(b)}|^2}-e\,\bm{E}(t),\label{eq1}\\
		m_i\,\frac{\mathrm{d}{\bm{v}}_\mathrm{i}^{(a)}(t)}{\mathrm{d}t}&=\frac{Z\,e^2}{4\pi\varepsilon_0}\sum_{\beta}\frac{\bm{n}_{a \beta}}{|\bm{r}_\mathrm{i}^{(a)}-\bm{r}_\mathrm{e}^{(\beta)}|^2}-\frac{Z^2\,e^2}{4\pi\varepsilon_0}\sum_{b\neq a}\frac{\bm{n}_{a b}}{|\bm{r}_\mathrm{i}^{(a)}-\bm{r}_\mathrm{i}^{(b)}|^2}+Z\,e\,\bm{E}(t),\label{eq12}
	\end{align}   
\end{widetext} where $m_e$ and $m_i$ are respectively the electron mass and the ion mass (only one population of ion is considered in this work), $Z$ is the charge number of an individual ion ($+Z\,e$ is the charge of an ion), $\bm{r}_\mathrm{e}^{(\alpha)}$ and $\bm{v}_\mathrm{e}^{(\alpha)}$ are the position and velocity of electrons (labeled with greek letters $\alpha$, $\beta$, $\cdots$) and $\bm{r}_\mathrm{i}^{(a)}$ and $\bm{v}_\mathrm{i}^{(a)}$ are the position and velocity of ions (labeled with roman letters for $a$, $b$, $\cdots$). The vectors $\bm{n}$ are unit vectors and $\bm{n}_{ab}$, for instance, is directed from particle $(a)$ to particle $(b)$. The external electric field, $\bm{E}(t)$, is uniform in space. The same assumption was made in Langdon's \cite{langdon1980} or Matte's \cite{matte1988} work because they were only concerned with  the local microscopic response of the plasma to the oscillating electric field. 

Since the pure Coulomb potential is not valid at very short range where the electronic structure of the ion should be taken into account, particles in CMDS interact with each other via a modified Coulomb potential. A soft core was added \citep{beutler1994} to the Coulomb potential in such a way that the interaction potential between two generic charges $q_1$ and $q_2$ is
\begin{align}
\phi (r) = \dfrac{q_1 \, q_{2}}{4 \, \pi \, \varepsilon_0 \, \sqrt{r_c^2 + r^2}},
\end{align}
where $r$ is the distance between $q_1$ and $q_2$ and $r_c$ is the approximate distance below which the pure Coulomb potential ceased to be meaningful. We chose $r_c=0.1\,\mathring{\mathrm{A}}$ and checked that our results do not depend upon the particular choice of $r_c$ provided it is small enough (see \citet{devriendt2022} for more details). Additionally, periodic boundary conditions were used to simulate an infinite uniform plasma and this is why LAMMPS' Particle-Particle Particle-Mesh (PPPM) algorithm was used to compute interactions with domain replicas.

Particular care was given to the production of the initial states of our CMDS. Equilibrium EVD and ion velocity distributions can be generated easily, since they are known to be \maxw at thermodynamic equilibrium. It is more complicated with the equilibrium spatial distributions of these particles. They were initially distributed at random in the simulation domain. Using a thermostat in LAMMPS, they were then left to relax until equilibrium was reached. This phase lasted for a few electron-ion collision times, until a steady state was established, which was monitored via the potential energy of the system, the ion and electron temperatures, the EVD, the quadrupole moment \cite{devriendt2022} and the three radial distribution functions (RDF) $g_{ee}(r)$, $g_{ei}(r)$ and $g_{ii}(r)$ (respectively for electron-electron, electron-ion and ion-ion RDF).

In order to encompass both subcritical and critical plasmas, where $n_c =9.05\cdot10^{21}$ cm$^{-3}$ for $\lambda=351$ nm, one initial state was generated with $n_e = 10^{20}$ cm$^{-3}$ (corresponding to $n_e\sim 0.01\,n_c$) and another one with $n_e =  10^{22}$ cm$^{-3}$ (corresponding to $n_e\sim n_c$). The choice of initial $T_e$ for both initialization was set by the choice of the plasma coupling parameter \citep{dimonte2008}
\begin{equation}
g =\dfrac{Z \, e^3 \, \sqrt{n_e}}{4 \, \pi \, (\varepsilon_0 \, \kB T_e) ^{3/2}},
\end{equation} that we fixed to the same value $g=0.05$ (which is $\ll 1$ corresponding to weakly coupled plasmas) for comparison purposes. That particular value is a compromise between plasma coupling weakness, for which one would advocate for small values of $g$, and not-too-high-numerical-cost, for which one would advocate for high values of $g$. The two initial conditions considered in this study are then:
\begin{align}
n_e &= 10^{20}\,\mathrm{cm}^{-3}, \, \kB T_e = 100\,\mathrm{eV}, \, Z=10,\label{eq:dist_plas1}\\ 
n_e &= 10^{22}\,\mathrm{cm}^{-3},\, \kB T_e = 500\,\mathrm{eV},\,Z=10.\label{eq:dist_plas2}	
\end{align} For both initial conditions, multiple initial states were generated, in order to reduce the effect of statistical fluctuations on our results.

In our CMDS, a time varying, spatially uniform, external electric field mimics the effect of a laser in a small simulation volume. It is periodic in time and the polarization is linear. The spatial uniformity of the electric field was the assumption in the MD simulations of \citet{pfalzner1998} and \citet{david2004}) and it is also the assumption of the theoretical derivation by Langdon \cite{langdon1980}. 

In a study dedicated to \maxw \mbox{$Z=1$} weakly coupled plasmas, the authors \cite{devriendt2022} devised a parametrized \maxw absorption model (PMAM) for electron-ion collision frequency in the IB heating process that encompasses most theoretical model in the literature. The model, with its six adjustable constants, $C_\mathrm{abs}$, $\eta$, $\varepsilon_\ell$, $C_\ell$, $\eta_\ell$ and $\delta$, reads
\begin{align}
\nu_{ei}^{IB}&= C_\mathrm{abs}\,\dfrac{4 \, \sqrt{2 \, \pi} \,e^4}{3 \, \sqrt{m_e}\, (4 \, \pi \, \epsilon_0)^2} \dfrac{n_e \, Z}{(\kB T_\mathrm{eff}(\eta))^{3/2}}\,\ln(\Lambda_{ei}^{IB}),\label{eq:dist_nu0}\\
\Lambda_{ei}^{IB}&=\left[\epsilon_\ell+C_\ell\,\frac{4\,\pi\,\varepsilon_0^{\frac{3}{2}}\,(\kB T_\mathrm{eff}(\eta_\ell))^{3/2}}{Z\,e^3\,\sqrt{n_e}}\right]\,\left(\frac{\omega_p}{\omega}\right)^{\bm{\delta}},\label{eq:dist_nueifin}
\end{align} where the effective temperature is defined as 
\begin{align}
	T_\mathrm{eff}(x)&=T_e+x\,\,m_e\,v_E^2/\kB\,.\label{eq:dist_teff}
\end{align} Its adjustable constants were fixed by comparison with CMDS results for $Z=1$ plasmas
\begin{align}
	C_\mathrm{abs}&=0.55,\label{cabs}\\
	\eta&=1/6,\\
	\epsilon_\ell&=1,\\
	C_\ell&=0.7,\\
	\eta_\ell&=1/6,\\
	\delta&=0.\label{delta}\
\end{align}

\par 

The heating rate of a uniform plasma submitted to an oscillating electric field of intensity $I$ is given by
\begin{align}
	\dfrac{\diff \kB T_e}{\diff t} = \dfrac{m_e}{3} \, \vosc^2 \, \nu_{ei}^{IB}.
\end{align}

\par

\begin{figure}[h!]
\centering
	\includegraphics[width=0.48\textwidth]{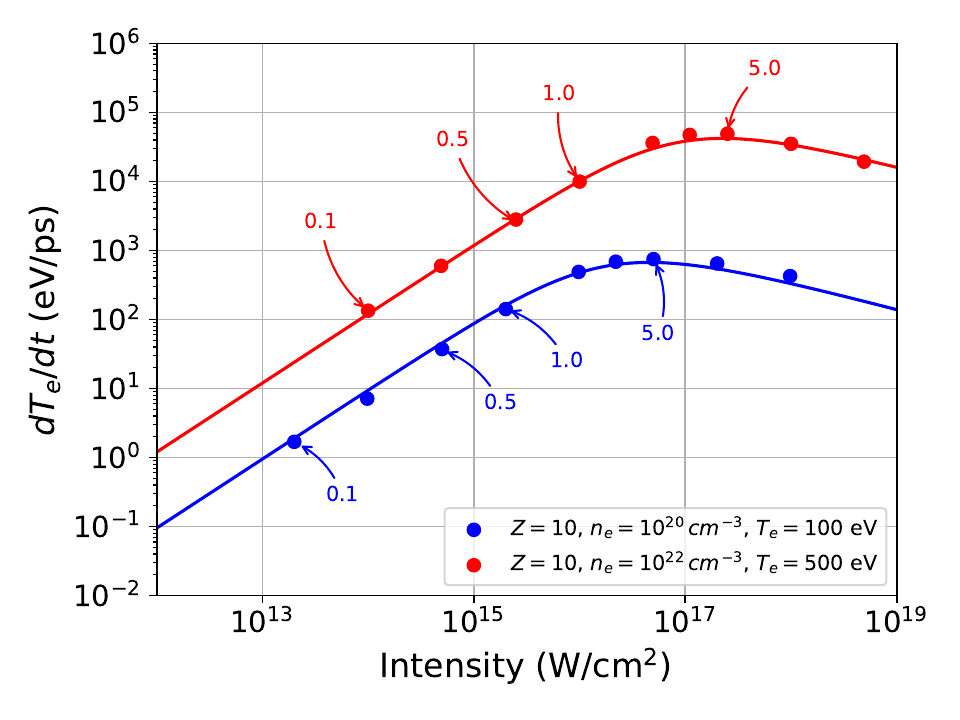}
	\caption{\label{fig:dist_simMDZ10}CMDS measures of the heating rate of plasmas in the conditions (\ref{eq:dist_plas1}) and (\ref{eq:dist_plas2}) at different intensities. The measurement is carried out at early time when EVDs have not yet transitioned to non-\maxw which is the hypothesis of the parametrized model developed in \cite{devriendt2022}. The coefficients of the model are those adjusted in \cite{devriendt2022} by CMDSs at $Z=1$, reported in (\ref{cabs})-(\ref{delta}), and one can see that the agreement is still excellent for $Z=10$. The colored numbers correspond to $\vosc$ in unit of initial $\vth$ for CMDS points indicated by the arrows. The best fit value is $C_\mathrm{abs}=0.45$ for the blue curve and $C_\mathrm{abs}=0.62$ for the red curve, still within one standard deviation ($\pm 0.07$) from the value $C_\mathrm{abs}=0.55$ found in \citet{devriendt2022} for the PMAM.}%
\end{figure}

In Fig. \ref{fig:dist_simMDZ10}, this parametrized model (solid lines), with constants adjusted for $Z=1$, is compared to CMDS results (points) of $Z=10$ plasmas initialized with condition (\ref{eq:dist_plas1}) or (\ref{eq:dist_plas2}). These points correspond to measurement of the heating rate in molecular dynamic simulations at early time. During this period, EVDs remain Maxwellian to a good approximation and can be compared to the parametrized model devised from various theoretical studies in the literature that hypothesized \maxw EVDs. Clearly, with the same parameters (\ref{cabs})-(\ref{delta}) fitted from $Z=1$ simulations, the agreement is still very good for $Z=10$ plasmas probed at early time. This last precision is of course very important since the remaining of this paper will deal with late time evolution of $Z=10$ plasmas where EVDs eventually turn non-\maxw.  

 In the fitted functions $k(\alpha)$ and $R(\alpha)$ provided by \citet{matte1988} (resp. order and reduction absorption coefficient), the values of $\alpha$ are not the initial values but the end-of-the-simulation value (when the self-similar behavior is achieved). The problem we observed in our CMDS is that the time it takes for the EVD to reach the self-similar regime increases with laser intensity. As a result, for a given CMDS duration, this limits the range of $\alpha$ where it is possible to observe that self-similar evolution of the EVD. Hence, our simulations were carried out in the range 
\begin{align}
	0.1 < \vosc/\vth^{0} < 5,
\end{align} where $\vth^0$ is the initial thermal velocity. Increasing the intensity further ($\vosc/\vth^{0}>5$) would have a high numerical cost because a smaller time-step would be needed to accurately capture high velocity electrons and the number of time iterations would jump drastically. On the contrary, reducing the intensity ($\vosc/\vth^{0}<0.1$) would come into conflict with signal to noise ratio in the CMDS because heating rate would be too small and buried under thermal fluctuations.

\section{Langdon effect impacts the instantaneous distribution in the oscillating frame}\label{sec::langdon}

In order to assess the effects of an oscillating electric field on the electron velocity distribution of a plasma, it is important to introduce two integrated distributions of practical interest for the description of the results presented in this work. The first one, the {\it isotropic distribution}, corresponds to the isotropic part of the 3D distribution. It is that isotropic distribution that was inferred by Langdon in its seminal paper \citep{langdon1980}. The second one, the {\it projected distribution}, is particularly useful when 3D distributions are anisotropic.

\subsection{Definition of various projection of the EVD}

In order to precisely define these two distributions, let us call the 3D distribution 
\begin{align}
	f_e(\bm{v})
\end{align} such that $f_e(\bm{v})\, \diff^3 \bm{v}$ corresponds to the number of electrons of velocity $\bm{v}$ within $\diff^3 \bm{v}$ in velocity space. That information can be extracted from any of our CMDS since the three components of the velocity of each individual particle, along with the three components of the position, are recorded at every time step.

{\bf The isotropic distribution} is computed from the 3D distribution $f_e(\bm{v})$, where the velocity vector can be recast as $\bm{v} = v\,\bm{\Omega}$, by fixing the modulus $v$ and averaging over all directions $\bm{\Omega}$:
\begin{align}
	f_e^0(v) = \dfrac{1}{4 \, \pi} \, \int f_e \left( \bm{v} \right) \, \diff^2\Omega.\label{eq:dist_isodist}
\end{align} It is used in Fokker-Planck codes, and is especially useful when the 3D distribution is close to isotropic which is the assumption in \citet{langdon1980}'s and \citet{matte1988}'s results. However, the laser electric field inherently introduces anisotropy in the 3D distribution as will be shown in \S\ref{sec:dist_highI}. It is thus appropriate to consider a last category of distributions.

{\bf Projected distributions} are obtained by integrating the 3D distribution over the two dimensions perpendicular to the axis of projection. As an example, the projected distribution on the $x$-axis requires to integrate over $y$ and $z$ such that 
\begin{align}
	f_x(v_x) = \int f_e(v_x, \, v_y, \, v_z) \, \diff v_y \, \diff v_z.\label{eq:dist_projdist}
\end{align} Since we only consider linear polarization in this study, two projections are of interest: along the direction of the electric field, and perpendicularly. Indeed, the projected distribution is expected to be the same along any of the directions perpendicular to the electric field. In the rest of this paper, quantities along the direction of the electric field will be referred to as parallel ($\varparallel$) and quantities referred to as perpendicular ($\perp$) are in a direction perpendicular to the electric field. 
 
%
%
%
Isotropic and projected distributions were extracted out of our CMDSs in the oscillating reference frame (also referred to as {\it centered velocity} in figures, where the quiver velocity is subtracted from particle velocities in the rest frame). It is in this oscillating reference frame that EVDs are closest to isotropic as observed from our CMDS (discussed in section \S\ref{sec:balescu}). The EVDs in the rest frame are averaged over one cycle of oscillation in such a way that

\begin{align}
	f_\mathrm{rest}(\bm{v}, t)=\frac{1}{T}\,\int_t^{t+T}\,f_\mathrm{osc}(\bm{v}-\bm{v}_\mathrm{osc}\,\sin(\omega\,\tau), \tau)\,\,\mathrm{d}\tau,
\end{align} where $T$ is one period of oscillation. If the EVD in the oscillating frame, $f_\mathrm{osc}$, is isotropic, the EVD averaged over one cycle of oscillation in the rest frame, $f_\mathrm{rest}$, is not unless $\vosc=0$.
 \begin{figure}[h!]
 \centering	
 \includegraphics[width=0.48\textwidth]{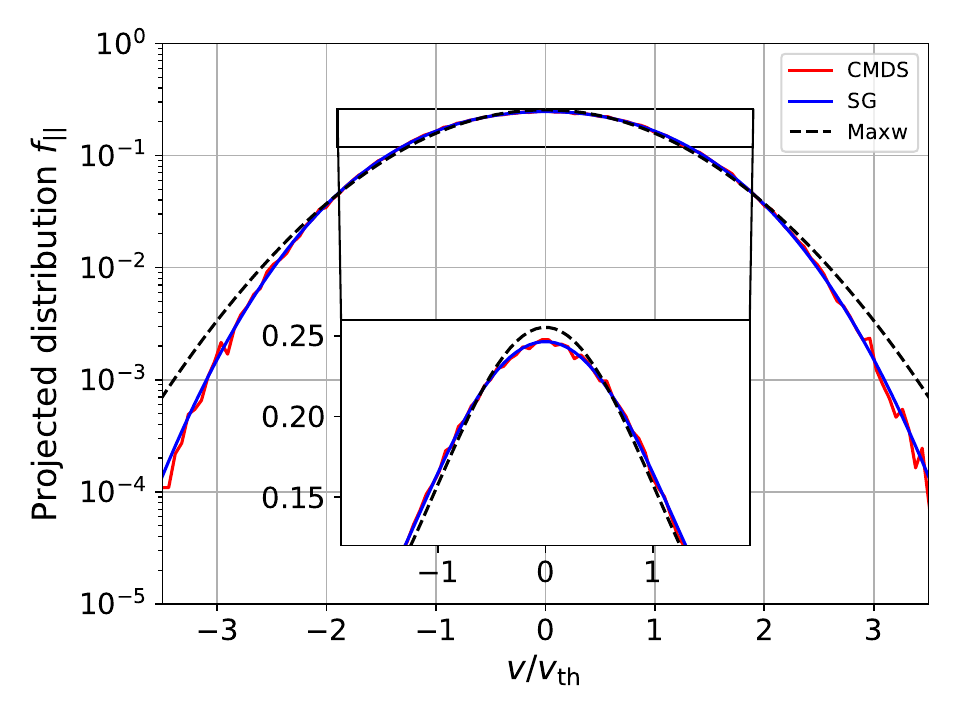}
 \includegraphics[width=0.48\textwidth]{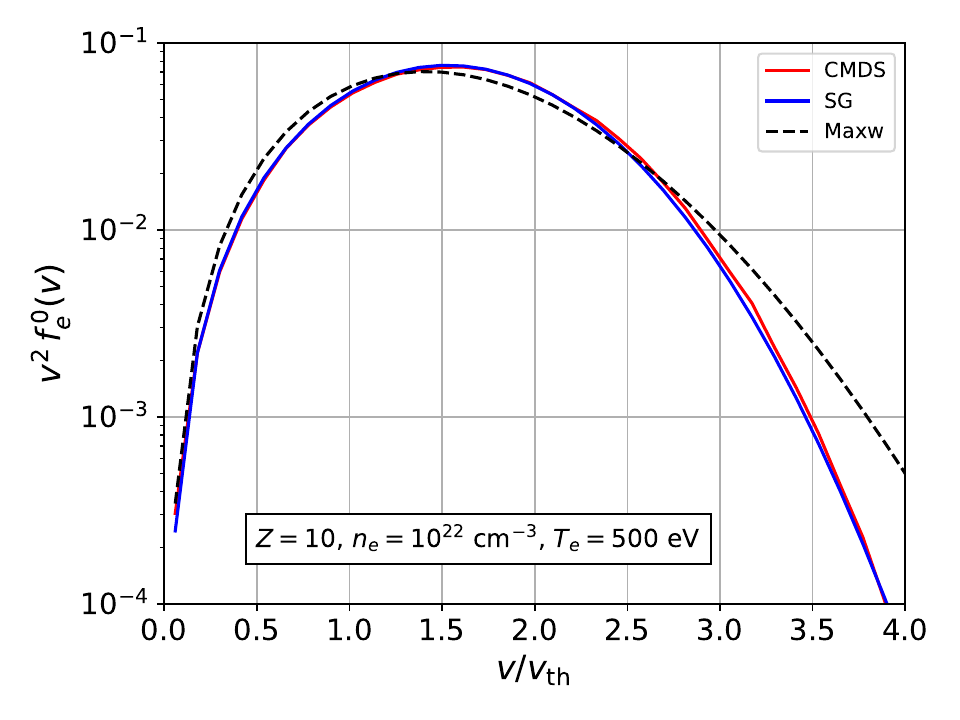}
  \caption{\label{fig:dist_langdon}Instantaneous projected (top) and isotropic (bottom) electron distribution function (in the oscillating frame) $f_{\varparallel}$ after 450 laser cycles. The simulation from which this result originates was set with $Z=10$, $T_e^0 = 500$ eV, $n_e = 10^{22}$ cm$^{-3}$ and $\vosc = 0.5 \,\vth^{(0)}$. As predicted by Langdon, the distribution function is closer to a \sg (SG) than to a \maxw (M). The best \sg fit corresponds to an order $k=2.69$.}%
 \end{figure}

\subsection{Langdon effect in CMDS}

In the conditions of the Langdon effect, $(\vosc/\vth)^2 \ll 1$ and $Z\,\vosc^2/\vth^2 \gg 1$, our CMDS display deformations of the instantaneous distribution function (in the oscillating frame). In Fig. \ref{fig:dist_langdon} the projected CMDS distribution on the parallel direction (to the laser polarization) $f_{\varparallel}$ is plotted and shown not to fit a projected \maxw (which is a \maxw). This is why we tested other theoretical distributions described in appendix \ref{sec:dist_distribExpr}. In Fig. \ref{fig:dist_langdon}, a projected \sg (\ref{eq:dist_projectedSG}), fits both its bulk and tail for $k=2.69$ at $\vosc/\vth = 0.5$ after 450 laser cycles, well in the self-similar regime (one recalls that $k$ is the \sg order of the 3D distribution). At smaller laser intensities ($\vosc/\vth = 0.1$), the projected CMDS distribution was also well fitted by a projected \sg, albeit with an order closer to 2. In both cases, the measured \sg order on our CMDS is less than the popular predictions by \citet{matte1988}. The discussion on this discrepancy is deferred to \S\ref{sec:dist_order} for it requires more elements that will be presented in the following sections. 

\par 

Notwithstanding this quantitative issue, the result presented in Fig. \ref{fig:dist_langdon} is, to our knowledge, the first observation in CMDS of non-\maxw effects on instantaneous EVDs in Langdon's conditions. 

\subsection{Our CMDSs seem to agree with Balescu : Langdon effect should be derived in the oscillating frame}\label{sec:balescu}

In the context of absorption by inverse bremsstrahlung, Langdon \cite{langdon1980} solved a Boltzmann equation which, for a uniform plasma submitted to a uniform external electric field, looks like
\begin{equation}
	\label{eq:dist_kinetic}
	\partial_t f_e - \dfrac{e}{m_e} \, \bm{E} \cdot \partial_{\bm{v}} f_e = C_{ee}(f_e) + C_{ei}(f_e),
\end{equation}
where $C_{ee}$ and $C_{ei}$ are the electron-electron and electron-ion collision operators (usually in the form of Landau collision operators \citep{langdon1980,jones1982}). Both collision terms are modeled and depend upon theoretical assumptions (as opposed to CMDS where collisions are inherently taken into account) on collision mechanism and on the influence of many-body effects, in particular, through Coulomb logarithms. 

\par 

Solutions of the Boltzmann equation (\ref{eq:dist_kinetic}) are developed in Legendre polynomials and only the first few terms are usually considered theoretically. \citet{langdon1980} retained only the slowly-varying (compared to the electric field) part of $f_e^0$ averaged over a laser period. The resulting equation is
\begin{equation}
	\partial_t f_e^0 = \dfrac{A \, \vosc ^2}{3} \, \dfrac{1}{v^2} \, \partial_v \dfrac{1}{v} \, \partial_v f_e^0,\label{eqlang}
\end{equation}
where $A = 2 \, \pi \, n_e \, Z \, e^4 \, \ln \Lambda /(m_e^2 \, (4 \, \pi \, \varepsilon_0)^2)$. Balescu \cite{balescu1982} derived the same equation by considering the oscillating reference frame of the electrons to separate the slow and fast varying parts of $f_e^0$. Balescu stated that {\it Langdon's derivation seems to be more sketchy. In particular, not enough care
	is taken in separating fast and slow processes in the distribution function: this
	can only be done in a simple way in the oscillating frame} and Weng \cite{weng2009} asserted that {\it the assumption of
	small anisotropy could only be well satisfied in the oscillating frame rather than in the
	rest frame}. This is not a cosmetic question for the result of Langdon theory is either valid in the rest frame (according to Langdon) or in the oscillating frame (according to Balescu). Langdon theory is valid in the reference frame where the EVD is closest to isotropic \cite{langdon1980} (this hypothesis is paramount to Langdon's derivation). Balescu \cite{balescu1982} showed that this derivation is more rigorously done in the oscillating frame and our CMDS seem to agree. The privileged reference frame where the EVD is closest to isotropic is the oscillating reference frame as observed in our CMDS (cf. Fig.\ref{fig:dist_anisotropicHeating}) as long as $\vosc/\vth\leq 1$. Outside this regime, when $\vosc/\vth> 1$, anisotropy develops (even in the oscillating frame) and Langdon theory ceased to be valid.

\section{Langdon effect at high laser intensity}
\label{sec:dist_highI}

\subsection{Anisotropy}

In agreement with previous works \citep{chichkov1992, pfalzner1992, ferrante2001}, our CMDSs show that the distribution isotropy is not preserved at increasing laser intensities.

\par

\begin{figure}[h!]
 \centering
 \includegraphics[width=0.48\textwidth]{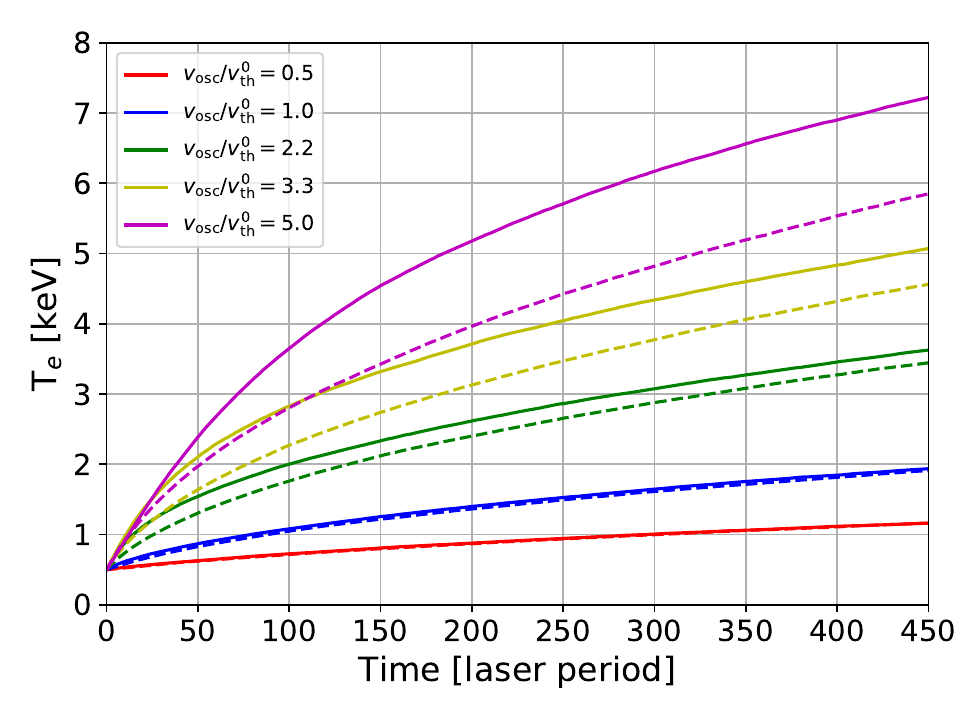}
 \caption{\label{fig:dist_anisotropicHeating} Evolution of perpendicular electronic temperature $\kB T_\perp$ (dashed lines) and parallel electronic temperature $\kB T_{\varparallel}$ (solid lines) for different initial values of $\vosc/\vth$ (these temperatures are calculated in the oscillating frame where the quiver velocity, $\bm{v}_\mathrm{osc}(t)$, as been subtracted from the velocities, in the rest frame, of all particles). The initial conditions correspond to (\ref{eq:dist_plas2}) where  $T_e(0) = 500$ eV and $n_e = 10^{22}$ cm$^{-3}$. When $\vosc\leq\vth$ temperature (and EVD in the oscillating frame) remains isotropic. When $\vosc > \vth$, anisotropy kicks in and the temperature in the direction parallel to the laser polarization rises faster than in the perpendicular directions.}%
 \end{figure}

In Fig. \ref{fig:dist_anisotropicHeating}, the evolution of the temperatures $\varparallel$ and $\perp$ to the polarization is plotted for multiple laser intensities, or equivalently for multiple initial values of $\vosc/\vth$.
In simulations where $\vosc > \vth$, it can be seen that the temperature in the direction of the polarization rises faster than in the other directions. The different temperatures are defined by $\kB T_{\varparallel} = m_e \, \langle{v^\prime}_{\varparallel}^2\rangle$ and $\kB T_\perp= \frac{1}{2}\,m_e \, \langle{v^\prime}_{\perp}^2\rangle$ where $\bm{v}^\prime$ is the fluctuating velocity of an electron around the ensemble averaged velocity (which is the quiver velocity $\bm{v}_\mathrm{osc}$) of the whole electron population and $\langle.\rangle$ is the ensemble average over that whole electron population. These temperatures can be accounted for in the distribution, while still preserving the \sg shape that we observed at lower laser intensities, with the single-order-anisotropic \sg distribution defined by (\ref{eq:dist_anisotropicSG}). This expression has been suggested by theoretical models \citep{porshnev1996} although the order was fixed and not variable as it is assumed here. With the appropriate temperatures, the $\varparallel$ and $\perp$ projections of the single-order-anisotropic \sg distribution (\ref{eq:dist_projectedSG}) closely fit the CMDS projected distributions $f_{\varparallel}$ and $f_\perp$.

\begin{figure}[h!]
	\centering
	\includegraphics[width=0.48\textwidth]{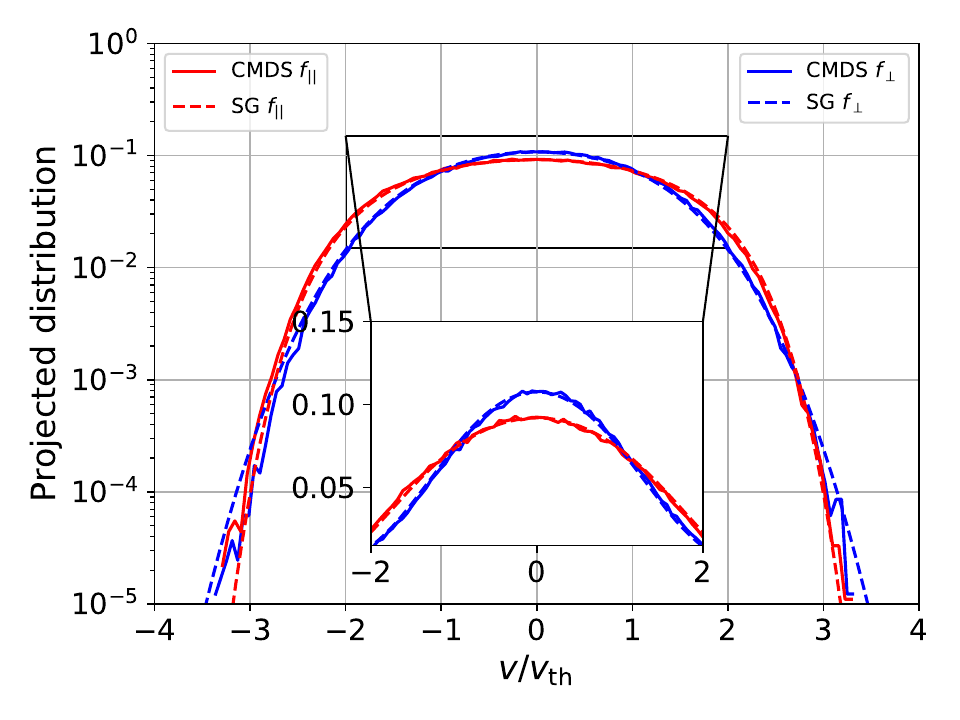}
	\includegraphics[width=0.48\textwidth]{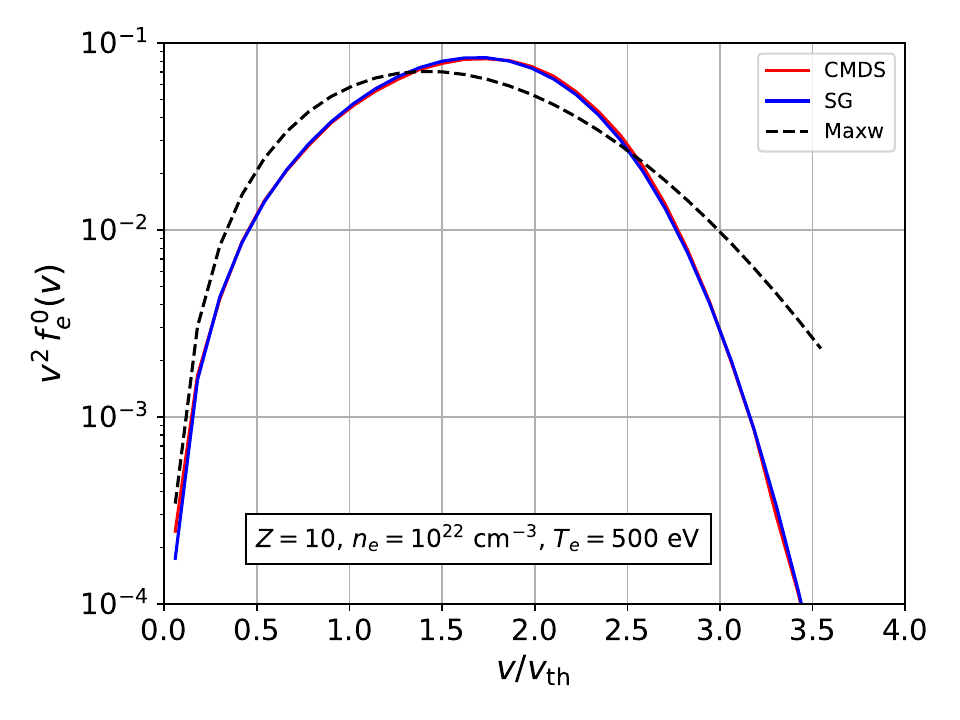}
	\caption{\label{fig:dist_distribAniso}\co Centered projected (top) and isotropic (bottom) distributions after 450 laser cycles. This simulation started with $Z=10$, $T_e^0 = 500$ eV, $n_e = 10^{22}$ cm$^{-3}$, and $\vosc/\vth^{0} = 5$. Both projected distributions (top) are well fitted by projected \sg, but with different order and thermal velocity. The best fit of $f_{\varparallel}$ with (\ref{eq:dist_projectedSG}) gives $v_{T{\varparallel}} = 3.78 \,\vth^{(0)}$ and $k_{\varparallel}=6.22$. In the perpendicular direction, $v_{T\perp} = 3.41 \,\vth^{(0)}$ and $k_\perp = 3.22$. This graph shows the distribution anisotropy both in temperature and order. The isotropic distribution (bottom) is well fitted by a \sg of order $k\approx4.0$.}%
\end{figure}

At even higher laser intensity ($\vosc \geq 2.2 \,\vth$), a single-order-anisotropic \sg distribution cannot fit both $f_{\varparallel}$ and $f_\perp$ simultaneously anymore. Indeed, as an illustration, the parallel and perpendicular instantaneous projected distributions of our CMDS data are plotted (solid line) on the same viewgraph in Fig. \ref{fig:dist_distribAniso}. It is a critical plasma ($n_e\approx n_c$) irradiated by a laser of intensity such that $\vosc=5\,\vth$. 
The perpendicular projected distribution $f_\perp$ is well-fitted by a projected \sg (\ref{eq:dist_projectedSG}) with $k\approx 3.5$ whereas the projected distribution along the direction of the polarization, $f_{\varparallel}$, is fitted by a projected \sg of order $k\approx 6$. A quantitative indicator of this anisotropy is the kurtosis of the projected distributions. It is defined by 
\begin{equation}
\textrm{Kurt}_\mu =  \frac{\left(\int v^4\,f_\mu(v)\,\mathrm{d}v\right)\left(\int f_\mu(v)\,\mathrm{d}v\right)}{\left(\int v^2\,f_\mu(v)\,\mathrm{d}v\right)^2}
\end{equation} where $\mu\in\{x, \, y, \, z\}$ is the projection direction. Any projected distribution of a \maxw distribution (which is itself a \maxw distribution) has a kurtosis of 3 regardless of the projection direction. For the single-order-anisotropic \sg distribution, defined in (\ref{eq:dist_anisotropicSG}), the kurtosis of its projected distribution is independent of the projected direction $\mu$, and is equal to:
\begin{equation}
\label{eq:dist_kurtSG}
\textrm{Kurt}_\mu =  \dfrac{9 \,\Gamma(3/k)\,\Gamma(7/k)}{5 \,\Gamma(5/k)^2}.
\end{equation} 

That kurtosis equals 3 when the order $k=2$ (for a pure \maxw). As one can notice, the kurtosis of a projected single-order-anisotropic \sg distribution is only a function of the order $k$ of the distribution even when the temperature is anisotropic. Since there is a one to one relation between the order $k$ of the EVD and the kurtosis of its projected distribution, (\ref{eq:dist_kurtSG}) can be reversed numerically and used to compute the order of the distribution, assuming the shape of the distribution is (\ref{eq:dist_anisotropicSG}). The order of the CMDS distribution can thus be computed in two ways: either by fitting the CMDS projected distribution to a projected \sg (\ref{eq:dist_projectedSG}) or by calculating the CMDS projected distribution kurtosis and using the inverse of relation (\ref{eq:dist_kurtSG}). The projection can be carried out along the electric field or perpendicular to it.

\par 

 Both methods lead to the same conclusion. The kurtosis analysis allows to easily assess the evolution of the order of the parallel and perpendicular projected distribution as time goes by. In Fig. \ref{fig:dist_kurtAniso}, the evolution of $\textrm{Kurt}_{\varparallel}$ and $\textrm{Kurt}_\perp$ is plotted for multiple laser intensities. The kurtosis can be read on the left vertical axis and the order of the projected \sg can be read on the right vertical axis. It is interesting to notice that for high intensities, the asymptotic $k_\perp$ does not depend upon intensity, at variance with the asymptotic $k_{\varparallel}$ that increases with intensity. Anisotropy seems inevitable as soon as $\vosc>\vth$.

\subsection{Supergaussian order can move away from the limits predicted by Langdon}
  
  \begin{figure}[h!]
  \centering
  \includegraphics[width=0.48\textwidth]{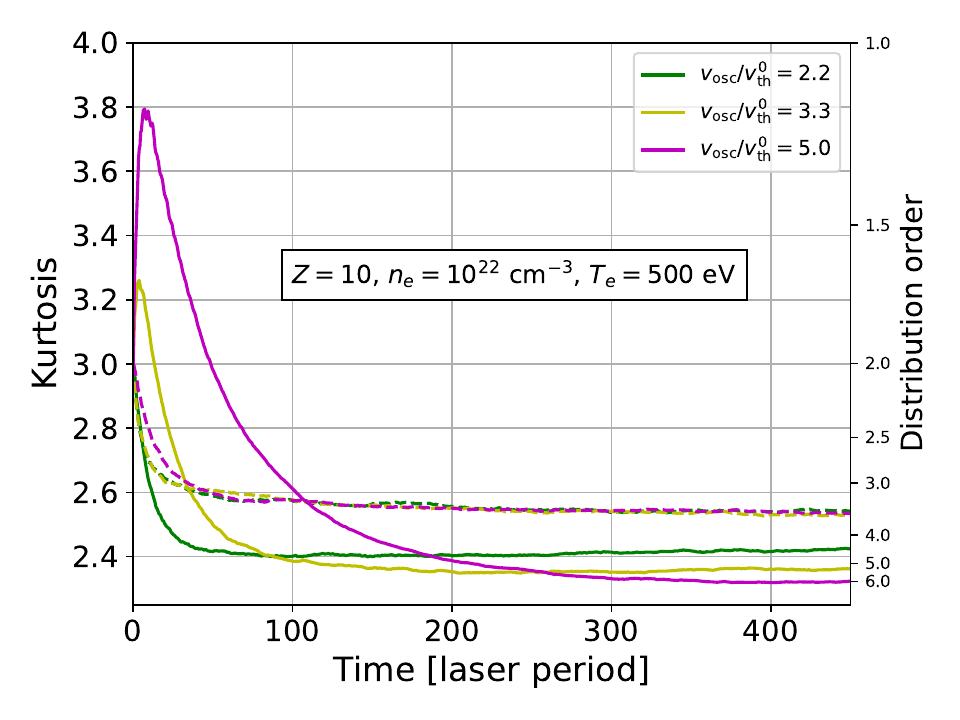}
	\caption{\label{fig:dist_kurtAniso}$\textrm{Kurt}_{\varparallel}$ (solid lines) and $\textrm{Kurt}_\perp$ (dashed lines) for different values of $\vosc/\vth$. The left vertical axis gives the value of the projected kurtosis ($\textrm{Kurt}_{\varparallel}$ or $\textrm{Kurt}_\perp$) and the right vertical axis is the corresponding SG order given by Eq. (\ref{eq:dist_kurtSG}). These simulations start from $\kB T_e^0 = 500$ eV and $n_e = 10^{22}$ cm$^{-3}$. When $\vosc > \vth$, the kurtosis becomes anisotropic. This is not compatible with a \sg distribution such as (\ref{eq:dist_anisotropicSG}).}%
\end{figure}  

After only a few laser cycles, for high intensities ($\vosc \geq  \vth$), the kurtosis can temporarily exceed 3 as can be seen in Fig.\ref{fig:dist_kurtAniso}. It implies that EVD projected \sg orders can be smaller than 2 before evolving toward a self-similar state. However, we have no reason to expect the distribution to fit a \sg shape before a self-similar state is reached, and so the computed order should be taken with caution.

\par 

\begin{figure}[h!]
   \centering
   \includegraphics[width=0.48\textwidth]{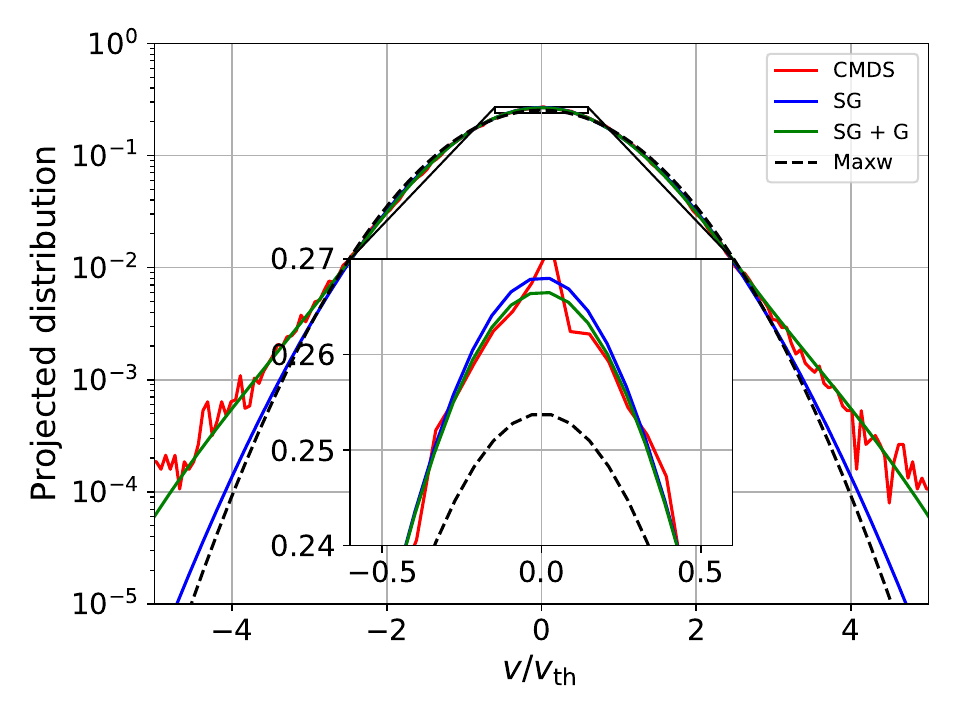}
   \includegraphics[width=0.48\textwidth]{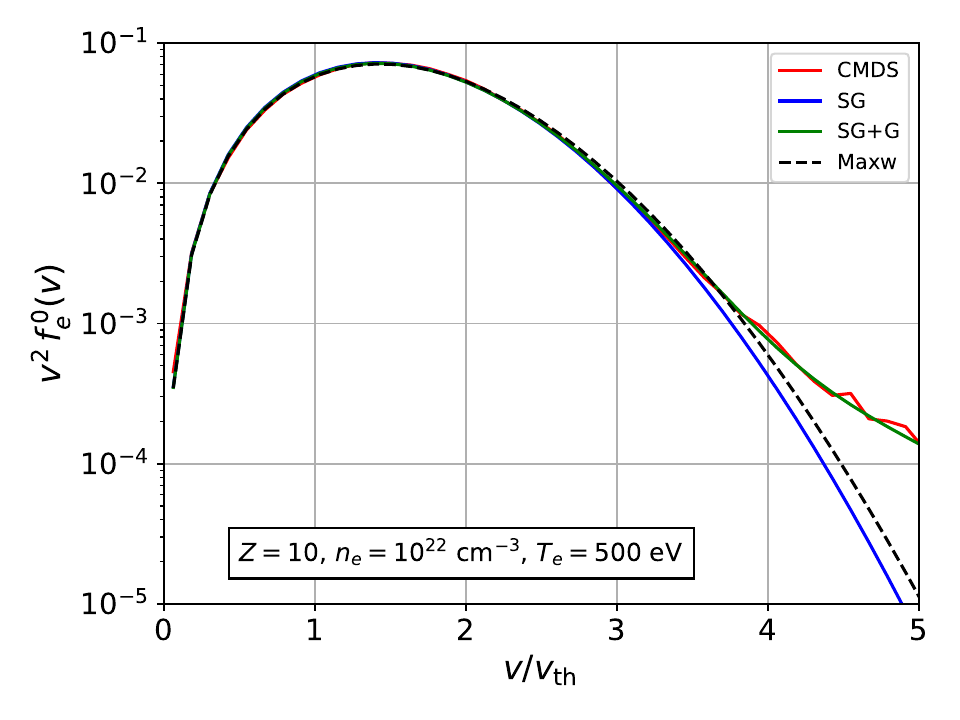}
  \caption{\label{fig:dist_histoTransient}Instantaneous projected $f_{\varparallel}$ (top) and isotropic $f^0_{e}$ (bottom) electron distribution function after 8.5 laser cycles. The data from CMDS is in red. It started with $T_e^0 = 500$ eV, $n_e = 10^{22}$ cm$^{-3}$ and $\vosc = 5 \, \vth^0$. The best \sg SG fit corresponds to $k=1.75$ and $\vth = 1.28 \, \vth^0$ and is in blue. For reference, the best \maxw with $\vth = 1.28 \, \vth^0$ is in dashed black. The best fit of a \sg plus a Gaussian (\ref{eq:dist_GplusSG}) corresponds to $A=-1.24$ (that is to say roughly 71$\%$ of SG and 29$\%$ of G), $\vth^G = 1.87 \, \vth^0$, $\vth^{SG} = 1.18 \, \vth^0$, and $k=1.97$ and is in green.}%
 \end{figure}

By examining the projected distribution at these instants in Fig. \ref{fig:dist_histoTransient}, one notices that the increase of kurtosis is mainly caused by a high-velocity tail in the EVD. Indeed, since the kurtosis is computed from the fourth power of the deviation to the mean, it is very sensitive to values far from the mean. The behaviour of the distribution agrees with the sum of two \maxw distributions with different temperatures in this transient regime. The small orders inferred from the kurtosis are evidence for the abundance of suprathermal electrons (strong tails in the EVD).

\begin{figure}[h!]
   \centering
   \includegraphics[width=0.48\textwidth]{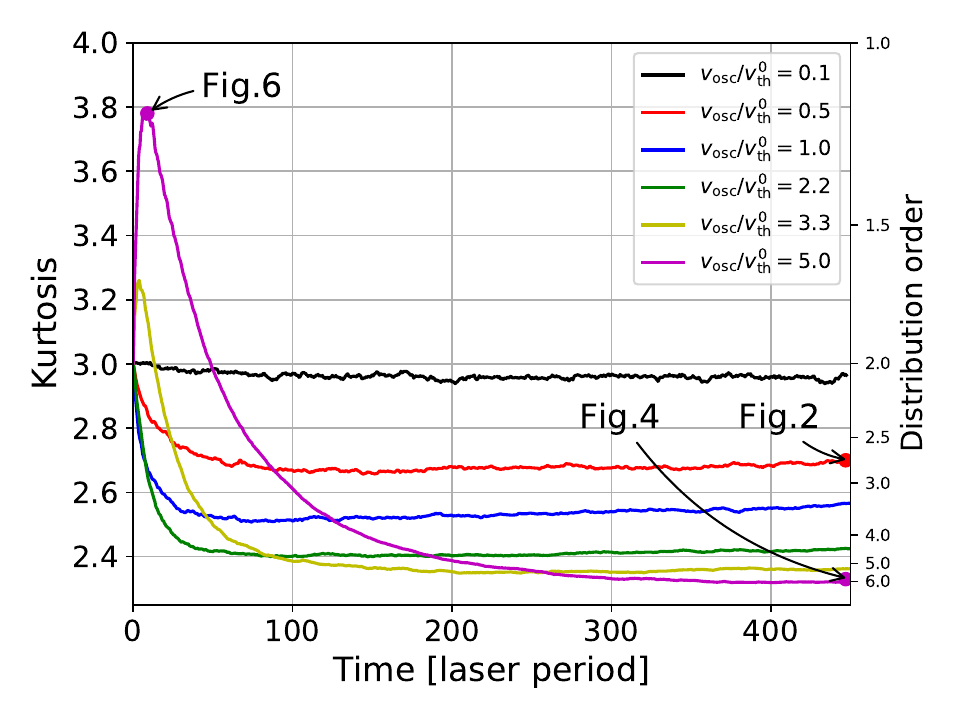}
 \caption{\label{fig:dist_kurtI}Evolution of $\textrm{Kurt}_{\varparallel}$ for multiple values of $\vosc/\vth$. On the right axis is the corresponding SG order given by (\ref{eq:dist_kurtSG}). These simulations start from $\kB T_e^0 = 500$ eV and $n_e = 10^{22}$ cm$^{-3}$. When $\vosc \geq 3.3 \,\vth$, a transient effect in which the kurtosis of the distribution becomes greater than 3 is observed. However, in the asymptotic behaviour of the kurtosis, which corresponds to a self-similar state of the distribution, the order seems to increase with increasing laser intensity.}%
 \end{figure}

In the self-similar regime, when the kurtosis leveled off in Fig.\ref{fig:dist_kurtI}, one can see that the asymptotic value of the kurtosis decreases with increasing laser intensity. This corresponds to increasing \sg orders. In particular, the \sg order can become higher than 5 unlike Langdon's prediction which is illustrated in Fig. \ref{fig:dist_distribAniso}. In fact, it is $k_{\varparallel}$ that can become greater than 5 whereas Langdon was specifically talking about the order of the isotropic distribution when predicting it could not exceed the value 5. We believe it is important to stress that the projected distribution can have higher orders than 5 and that $k_{\varparallel}$ is always above the order of the isotropic distribution in the self-similar regime as will be discussed in Fig.\ref{fig:dist_order2eta}.

In a nutshell, we observed three kinds of anisotropy in the distribution at high $\vosc$: the coherent oscillating movement of the electrons which causes anisotropy in the rest frame, the temperature anisotropy, and the anisotropy of the distribution order.

\subsection{Observation of \sg distributions with \maxw tails}\label{sec:dist_tail}

  \begin{figure}[h!]
	\centering
	\includegraphics[width=0.48\textwidth]{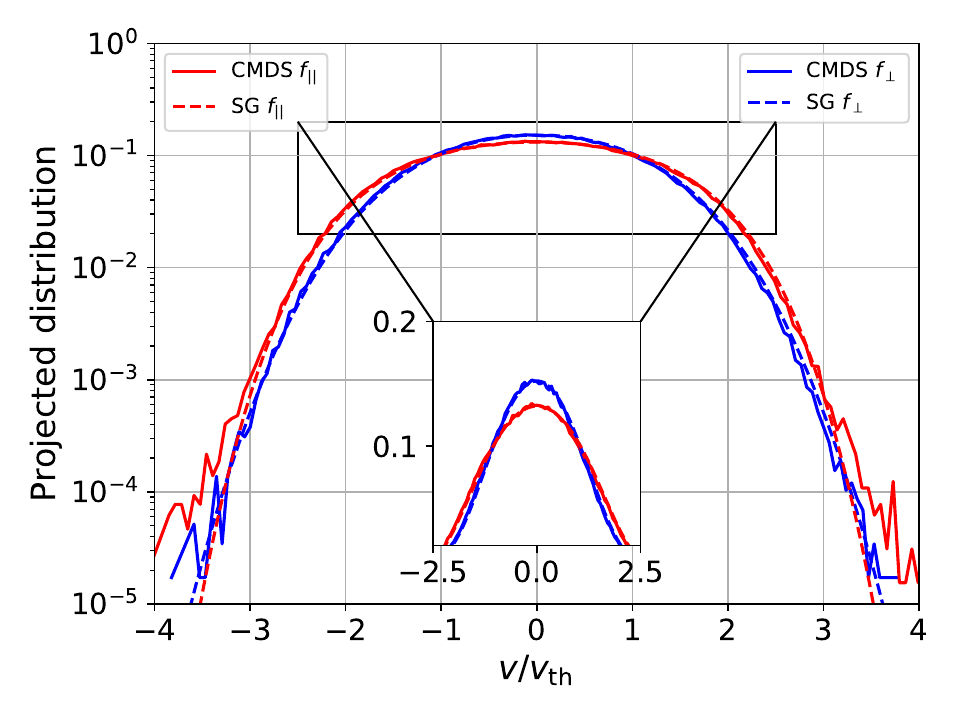}
	\includegraphics[width=0.48\textwidth]{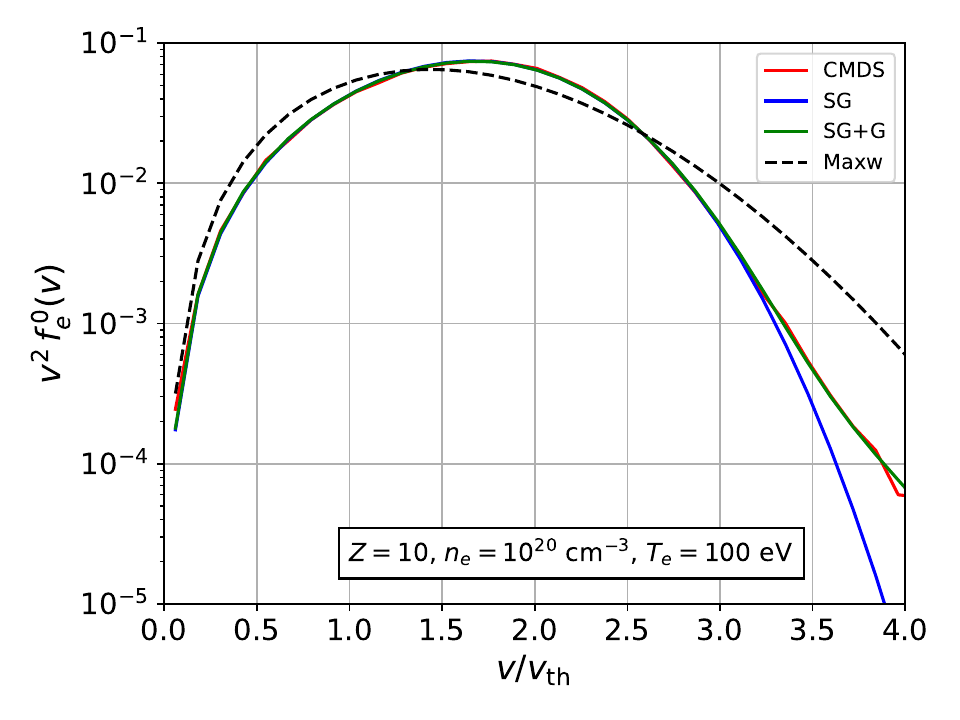}
	\caption{\label{fig:dist_distribIso}Instantaneous projected (top) and isotropic (bottom) distribution after 2963 laser cycles (solid red line). The best \sg fit is in solid blue, and the best fit of the sum of a \maxw and \sg (\ref{eq:dist_GplusSG}) is in solid green. This simulation started from $\kB T_e^0 = 100$ eV and $n_e = 10^{20}$ cm$^{-3}$, and $\vosc/\vth^0 = 3.3$. The low velocity part of the distribution is very well fitted by a \sg distribution. However, the tail of the distribution behaves like a \maxw. The \maxw in dashed black correspond to the one with the same total electron kinetic energy as the CMDS data. The isotropic \sg order is measured here to be $k=3.53$. The parallel projected distribution order is $k_{\varparallel} = 4.43$ and the perpendicular projected distribution order is $k_\perp=3.08$. With the notation of \cite{milder2021}, our CMDS results were fitted with formula (\ref{eq:dist_GplusSG}), and it was found that $A_1=0.0345$, $x_1=1.94$, $k=3.53$, $A_2=-5.93$ and $x_2=1.53$ which is in qualitative agreement with results for shot 95830 in \cite{milder2021}.}%
\end{figure} 

   \begin{figure*}[ht]
	\centering
	\includegraphics[width=0.48\textwidth]{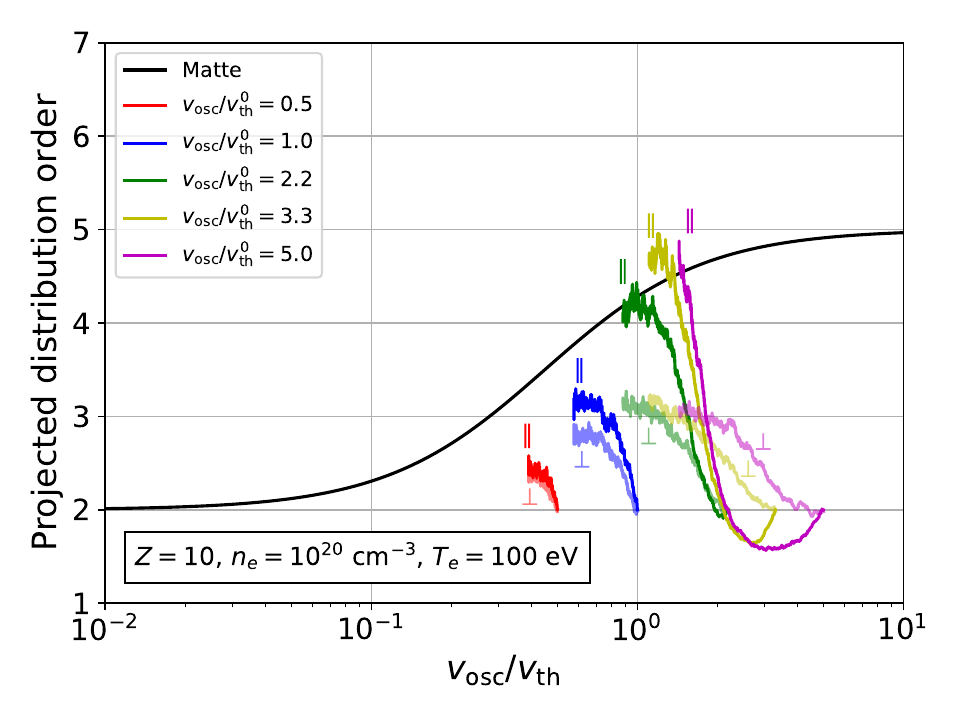}  \includegraphics[width=0.48\textwidth]{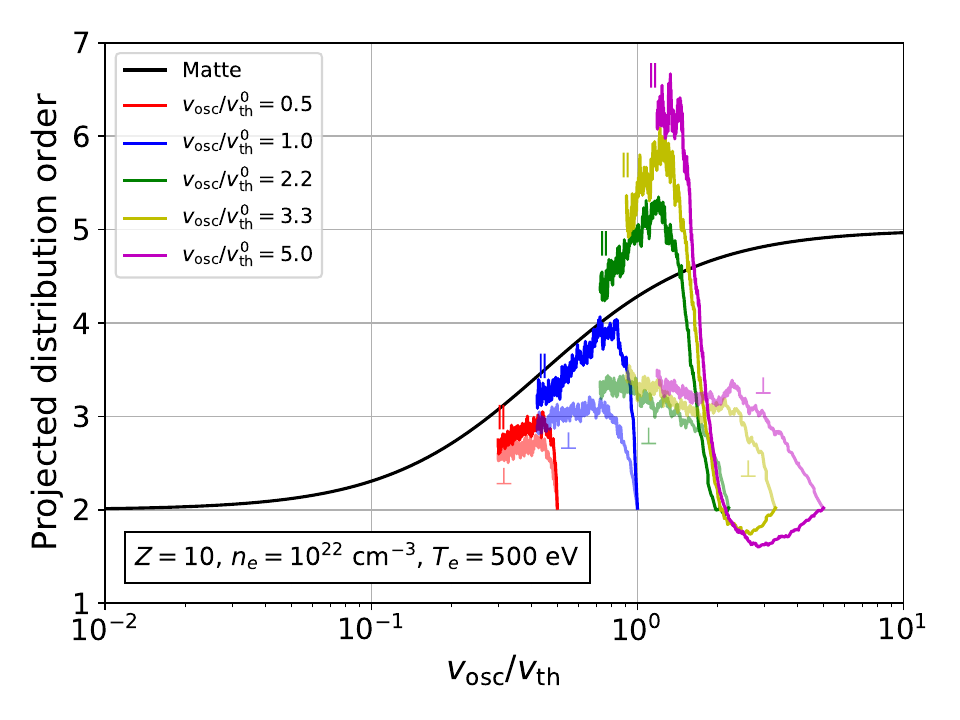} \\
	\includegraphics[width=0.48\textwidth]{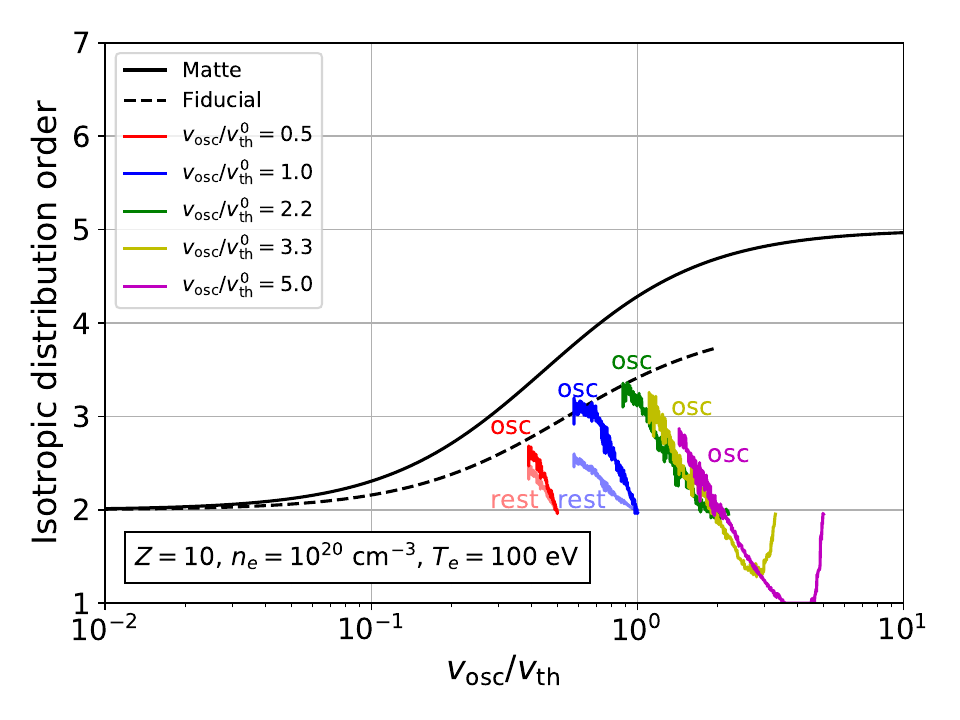} \includegraphics[width=0.48\textwidth]{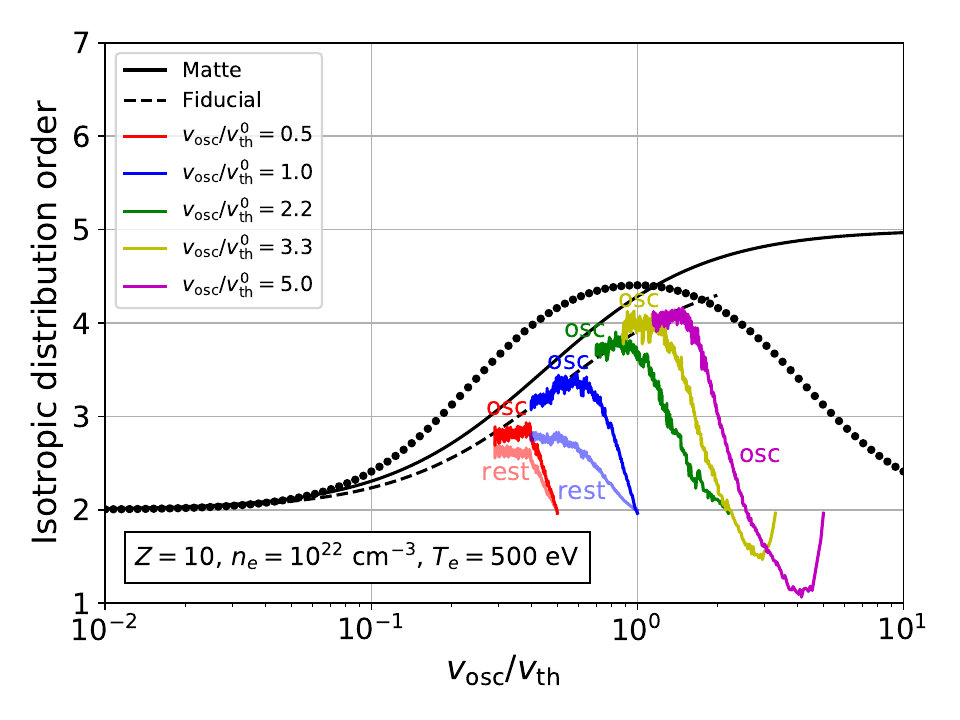}
	\caption{\label{fig:dist_order2eta} \sg order as a function of $\vosc/\vth$ for multiple laser intensities. The top viewgraphs correspond to projected-distribution-order in the oscillating reference frame and the bottom viewgraphs to isotropic distribution order with respect to $\vosc/\vth$ in the oscillating reference frame ($\mathrm{osc}$) or in the rest frame ($\mathrm{rest}$). Colors allow to distinguish initial values of $\vosc/\vth$. One color corresponds to one CMDS with a given laser intensity kept constant during the simulation. The time evolution on each of these curves goes from right to left because, as time goes by, $T_e$ keeps increasing due to the IB heating and $\vosc/\vth$, which is proportional to $\sqrt{I/\kB \,T_e\,\omega^2}$, decreases accordingly. The black solid line on each viewgraph is Matte's\cite{matte1988} prediction and the black dotted line in the bottom right figure corresponds to Weng's\cite{weng2009} equation (17). These simulations begin either with plasma state (\ref{eq:dist_plas1}), in right viewgraphs, or with plasma state (\ref{eq:dist_plas2}), in left viewgraphs, and with a \maxw EVD (order $k=2$). (top) Pale colored curves correspond to perpendicular ($\perp$) projected-distributions orders and dark colored curves of the same color correspond to parallel ($\varparallel$) projected-distribution orders. (bottom) Pale colored curves ("rest") correspond to isotropic distributions orders in the rest frame and dark colored curves ("osc") correspond to isotropic distributions orders in the oscillating frame. The fiducial (dashed black) in the left viewgraph is $k=2+2/(1+2.2/\alpha^{0.724})$ and the fiducial in the right viewgraph is $k=2+2.6/(1+1.9/\alpha^{0.724})$.}%
\end{figure*}

%

A model was proposed by \citet{fourkal2001} in situations where the non-\maxw evolution leads to EVDs that are superposition of a \sg distribution with a \maxw distribution.
The low-velocity part of the distribution (with respect to the thermal velocity) is expected to be distorted by the Langdon effect thus becoming \sg, but the dynamics of the high-velocity part of the distribution are dominated by electron-electron collisions.
These collisions tend to relax the distribution towards a Maxwellian, in the electrons oscillating frame. Taking both of these effects into account, the distribution has the appearance of a \sg with a \maxw tail. This theoretical model was shown to be in agreement with experimental measurements of the isotropic distribution function \citep{milder2021}. 

\par 

Although the exact experimental conditions \cite{milder2021} were too numerically costly to replicate in our CMDS, we carried out simulations in similar conditions of electronic density (in the experiments $n_e \approx 4\times 10^{19}$ cm$^{-3}$ and in our CMDS $n_e = 10^{20}$ cm$^{-3}$), electronic temperature ($\kB T_e = 0.5 - 1$ keV in the experiments, while our simulations start with $\kB T_e^0 = 100$ eV, but the electrons heats up to  $900$ eV in our CMDS), and with ionization $Z=10$ (instead of $Z\approx 25$ given by the atomic physics code SAPHyR \cite{poujade2021} at $n_e = 4\times 10^{19}$ and $T_e=$ 1 keV). Fig.\ref{fig:dist_distribIso} shows that, in our CMDS, we also observe isotropic distributions that are \sg at low velocities with \maxw tails at high velocity (with respect to $\vth$). The EVD is displayed after 2963 laser cycles which corresponds to $\vosc/\vth=1.2$ (as can be seen in Fig.\ref{fig:dist_order2eta}, the end of the simulation of $\vosc/v^0_\mathrm{th}=3.3$ for $Z=10$, $n_e=10^{20}$ cm$^{-3}$ and $T_e=100$ eV corresponds to $\vosc/\vth=1.2$). Fig.\ref{fig:dist_distribIso} also shows that the transition from \sg to \maxw appears around $v=3\,\vth$ as observed experimentally in Fig.5 of \cite{milder2021}.

\par 

Qualitatively, CMDS provide the same behavior as observed experimentally and we have observed that a \sg with \maxw tail EVDs only show up, in our CMDS, when $n_e<n_c$.

 
\section{The order of the distribution as a function of $\vosc/\vth$}\label{sec:dist_order}

Fokker-Planck simulations were carried out by \mbox{Matte et. al.} \cite{matte1988} in order to generalize the work of Langdon by taking into account both electron-electron and electron-ion collision terms. An empirical fit on the order of the isotropic distribution with respect to $\alpha = Z \, \vosc^2/\vth^2$ was proposed by Matte
\begin{align}
\label{eq:dist_matteOrder}
k(\alpha) = 2 + \dfrac{3}{1+\dfrac{1.66}{\alpha^{0.724}}},
\end{align}
along with the reduction in laser absorption due to the distortion of the distribution. As pointed out by Balescu \cite{balescu1982}, Langdon's conclusion and therefore Matte's, are valid in the reference frame where the EVD is closest to isotropic. Balescu showed that it is the oscillating frame and our CMDS agree. Indeed, in Fig.\ref{fig:dist_order2eta}, the bottom right viewgraph shows that the isotropic distribution order in the oscillating frame ("osc") in the self similar regime closely follow Matte's fit in the self-similar regime at variance with the isotropic distribution order in the rest frame ("rest").

\par 

One can produce fits for $k(\alpha)$ (in the oscillating frame) out of our CMDS and compare them with those obtained in \citet{matte1988} by FP simulations in the case of uniform plasmas. As mentioned by these authors, in the absence of sink mechanism such as conduction (which is the case for FP simulations of uniform plasmas and also for our CMDS) the value of $\alpha$, defined in (\ref{eq:dist_defAlpha}), is not constant and decreases as the plasma electron temperature ($T_e$) increases due to the laser heating at constant intensity.
\par

In Fig.\ref{fig:dist_order2eta}, the top viewgraphs correspond to the evolution of the \sg order of the $\varparallel$ and $\perp$ projected distributions and the bottom viewgraphs to the evolution of isotropic distribution. Each curve in each viewgraph corresponds to the evolution of the \sg order of a single CMDS starting as a Maxwellian (with order 2 on the vertical axis). The initial value of $\vosc/\vth$ can be read on the horizontal axis of order 2. On each curve, the laser intensity is kept constant and $\vosc/\vth$ decreases as time goes by because electrons heats up. Therefore, on each curve, the time evolution is from right to left. Simulations do not reach a steady state since the order keeps evolving as the simulation proceed. As a reference, Matte's formula, given in (\ref{eq:dist_matteOrder}) for the order of the distribution, is plotted in solid black. 

One can see in the top right viewgraph of Fig.\ref{fig:dist_order2eta} that parallel-projected-distributions-order can overshoot Matte's prediction unlike perpendicular-projected-distribution-order that seems to remain well under at all time. There seems to be a change in behaviour at early time when $\vosc/\vth\geq 3.3$. Above this threshold, the parallel-projected-distribution-order seems to dive first with orders less than 2 (going from left to right), which indicates the creation of a supra-thermal electron tail as explained in \S\ref{sec:dist_tail}, before rising rapidly and reaching the self-similar regime. There seems to be a dependency of $k_{\varparallel}$ upon the initial electronic density, in addition to the obvious dependency upon $\vosc/\vth$, when comparing the top right and top left viewgraphs in Fig. \ref{fig:dist_order2eta}. The plasma coupling parameter is the same initially for both initial conditions ($g=0.05$). On the contrary, $k_{\perp}$, as a function of $\vosc/\vth$, seems much less affected by the electronic density.   

In the bottom right viewgraphs of Fig.\ref{fig:dist_order2eta}, when $n_e\approx n_c\approx 10^{22}$ cm$^{-3}$, one can clearly see that CMDSs do not reach a steady state. After a transient regime, the evolution of the isotropic distribution order ends up following the curve by Matte\cite{matte1988} to within 10 $\%$, irrespective of the initial value of $\vosc/\vth$. This is the self-similar regime predicted by Langdon\cite{langdon1980}. 

\par 

In the bottom left viewgraphs of Fig.\ref{fig:dist_order2eta}, when $n_e\approx 0.01\,n_c\approx 10^{20}$ cm$^{-3}$, simulations did not go as far (more time consuming), but one can sense on the blue and green curve an inflection towards a self similar regime that would evolve along a curve (represented in dashed black) at lower isotropic distribution orders than Matte's fit.

The discrepancy between our observations and the prediction of \citet{matte1988} may be traced back to the fact that Matte {\it et. al.} used Fokker-Planck simulations where EVDs shapes are constrained by the limited development in Legendre polynomials in FP codes. Another sticking point concerns collisions that have to be modeled. In particular, the Coulomb logarithm $\ln \Lambda$ that appears in the electron-ion collision frequency, in (\ref{eq:dist_nu0}) for instance, has to be provided as an input to FP simulations, which is not the case in CMDS.

\section{Conclusion}

We observed the distortion of the EVD in moderate-Z plasmas for a wide range of laser intensities. At low laser intensity, we were able to observe the distortion predicted by \citet{langdon1980} on the instantaneous EVD (in the oscillating frame), which had yet to be done in CMDS with $\vosc < \vth$ and $Z>1$. Our CMDSs show that the reference frame where EVDs are closest to isotropic is the oscillating frame, as emphasized by Balescu \cite{balescu1982}, and it is therefore in this referential that Langdon's proof should be correct. At higher laser intensity, we observed and characterized the distribution of the EVD. In particular, in our CMDSs, the projected distribution along the direction of the laser polarization was observed to reach a \sg order higher than 5 while the isotropic distribution order always stayed bellow that value which was the highest order possible predicted by Langdon for the isotropic distribution. We observed \sg distributions with \maxw tail in situations similar to experimental conditions \citep{milder2021}. Finally, we showed that the behaviour of the isotropic part of the \sg is qualitatively similar to the predictions of \citet{matte1988}, but with a \sg order slightly smaller from a quantitative point of view.

\appendix

\section{Expressions for various useful \sg distributions and their projection}
\label{sec:dist_distribExpr}

The non-\maxw expression of the EVD, first devised by \citet{langdon1980} and also used by \citet{jones1982}, \citet{matte1988} and many others, is that of an isotropic \sg (ISG) given by 
\begin{equation}
\label{eq:dist_isotropicSG}
f_e^{\scriptscriptstyle \mathrm{ISG}}(\bm{v}) =  \dfrac{k \, \Gamma(5/k)^{3/2}}{4 \, \pi \, \vth^3 \, \Gamma(3/k)^{5/2} \, 3^{3/2}} \exp \left[ -\left(\dfrac{v}{v_{\scriptscriptstyle \mathrm{I}}}\right)^k \right],
\end{equation}
where the velocity $v=(v_x^2+v_y^2+v_z^2)^{1/2}$, the width $v_{\scriptscriptstyle \mathrm{I}} = \vth\times (3 \, \Gamma(3/k)/\Gamma(5/k))^{1/2}$ and $k$ is the order of the distribution. In the theoretical derivation of this expression by Langdon, isotropy was hypothesized. But as shown in \S\ref{sec:dist_highI} of the present article, our CMDS show that EVDs can become anisotropic for different reasons.

At small to moderate intensities, CMDS illustrate that different temperatures can build up in both parallel and perpendicular directions while keeping a common order $k$ in all directions (the reference being the direction of polarization). Therefore a single-order-anisotropic \sg (SOA) distribution written as 
\begin{equation}
\label{eq:dist_anisotropicSG}
f_e^{\scriptscriptstyle \mathrm{SOA}}(\bm{v}) =C\, \exp   \left[- \left(\dfrac{v_x^2}{{v^2_{\scriptscriptstyle \mathrm{S}}}_x} + \dfrac{v_y^2}{{v^2_{\scriptscriptstyle \mathrm{S}}}_y} + \dfrac{v_z^2}{{v^2_{\scriptscriptstyle \mathrm{S}}}_z} \right)^{k/2}  \right],
\end{equation} is enough to describe such non-\maxw EVDs. In (\ref{eq:dist_anisotropicSG}), $C$ is a normalization constant and the width ${v^2_{\scriptscriptstyle \mathrm{S}}}_\mu = {\vth}_\mu\times (3 \, \Gamma(3/k)/\Gamma(5/k))^{1/2}$ with $\mu\in\{x,\,y,\,z\}$. When projected on the $\mu$-axis, in the sense of (\ref{eq:dist_projdist}), this last distribution becomes
\begin{equation}
\label{eq:dist_projectedSG}
f_\mu^{\scriptscriptstyle \mathrm{SOA}} (v_\mu) = \dfrac{\sqrt{\Gamma(5/k)}}{2 \sqrt{3} \vth \Gamma(3/k)^{3/2}} \,\,\gamma \left( \dfrac{2}{k}, \left( \dfrac{v_\mu}{{v_{\scriptscriptstyle \mathrm{S}}}_\mu} \right)^k \right),
\end{equation}
where $\gamma(a, z) = \int_0^z t^{a-1} \exp (-t) \diff t$ is the incomplete lower gamma function. In the case of a 3D \maxw distribution, for which $k=2$ in (\ref{eq:dist_anisotropicSG}), the projection is trivially a \maxw with the same temperature as that along the projected axis. 

\par 

At high intensities, our CMDS shows that even the order is anisotropic (see \mbox{Fig.\ref{fig:dist_order2eta}} for instance). A two-order-anisotropic \sg (TOA) distribution of the form
\begin{equation}
	\label{eq:dist_manyanisotropicSG}
	f_e^{\scriptscriptstyle \mathrm{TOA}}(\bm{v}) =C\, \exp   \left[-\! \left(\frac{v_x^2}{{v_{\scriptscriptstyle\mathrm{I}}}^2_{\scriptscriptstyle\parallel}}\right)^{k_\parallel/2} \!\!\!- \left(\frac{v_y^2+v_z^2}{{v_{\scriptscriptstyle\mathrm{I}}}^2_{\scriptscriptstyle\perp}}\right)^{k_\perp/2}   \right],
\end{equation} is enough to describe such non-\maxw EVDs. Here, we have supposed that the polarization is along the $x$ axis. 

\par 

Finally, using particle-in-cell simulations, \citet{fourkal2001} found that the non-\maxw behaviour of EVDs in a plasma submitted to an intense monochromatic radiation (laser) had one more feature that was not described by Langdon's seminal paper \citep{langdon1980}. Although the bulk of the EVD is well described by a \sg, its tail (for $v\gg \vth$) was found to match a \maxw. This is a characteristic that we were able to observe in our CMDS and this is the reason why it is listed in this section. It is represented by a \sg-\maxw tail (SGM) function of the form (same notation as in \cite{milder2021})
\begin{align}
\label{eq:dist_GplusSG}
f_e^{\scriptscriptstyle \mathrm{SGM}}(v) =& A_1\, \exp\left[-\left(\frac{v}{x_1}\right)^k\right] \nonumber\\
&+\exp[A_2]\, \exp\left[-\left(\frac{v}{x_2}\right)^2\right],
\end{align} where $A_1$ and $\exp[A_2]$ are parameters that characterize, respectively, the weight of the \sg and of the \maxw distribution. In \cite{milder2021}, normalization is such that 
\begin{align}
	\int_0^{+\infty}\,4\,\pi\,v^2\,f_e^{\scriptscriptstyle \mathrm{SGM}}(v)\,\mathrm{d}v=1
\end{align}

\section*{Data Availability}

The data that support the findings of this study are available
from the corresponding author upon reasonable request.

\section*{References}
\bibliography{biblio}
\end{document}